# Self-Organized Networks and Lattice Effects in High Temperature Superconductors I: Lattice Softening

J. C. Phillips

Dept. of Physics and Astronomy, Rutgers University, Piscataway, N. J., 08854-8019

**Abstract**

The self-organized dopant percolative filamentary model, entirely orbital in character (no fictive spins), explains chemical trends in superconductive transition temperatures $T_c$, assuming that Cooper pairs are formed near dopants because attractive electron-phonon interactions outweigh repulsive Coulomb interactions. According to rules previously used successfully for network glasses, the host networks are marginally stable mechanically. The high $T_c$'s are caused by softening of the host network, enormously enhanced by large electron-phonon interactions at interlayer dopants for states near the Fermi energy. Background (in)homogeneities (pseudogap regions) produce novel percolative features in phase diagrams.

## 1. Introduction

High temperature cuprate superconductivity may be the most complex phenomenon known in inorganic materials. It has been the subject of more than 65,000 papers, and a large number of theoretical models have attempted to explain the many counter-intuitive phenomena observed. Many of the theoretical papers contain elaborate and ingenious mathematical models whose relation to experiment is vague. In this paper topological methods, based on the author's earlier theories of the glassy behavior of dopants in the cuprates [1], are used to discuss several anomalies in detail, with emphasis on the key role played by the connectivity of the internal dopant structure. The theory emphasizes qualitative trends, as experience has shown that that the complexity of these materials may well preclude quantitative treatments of the kind that worked so well for simpler superconductors, such as $MgB_2$ [2].

Scanning tunneling microscopy (STM) has revealed a strongly disordered, patchy (~ 3 nm) pattern of gap inhomogeneities in $Bi_2Sr_2CaCu_2O_{8+x}$ (BSCCO) [3-5], with patterns strongly dependent on the concentration x of interstitial oxygen dopants $O_x$. These patterns are the result of projecting 3-dimensional structure onto a 2-dimensional field of view, which still leaves many aspects of the true structure unknown. Continuum methods (including Fourier transforms) explain only a very small part of these patterns, with the remainder being left as mysterious "dark matter". The dopant network model explains many more aspects, all within a unified framework. The complex, partially hidden and strongly disordered structure of the cuprates appears to be essential to their unparalleled properties as HTSC: for example, the development of superlattice ordering at commensurate doping concentrations (the 1/8 phase of LSCO) greatly reduces $T_c$, probably to zero when fully developed.. Experience with exponentially complex molecular and network glasses has shown that no single polynomial mechanism describable by mean field theory can provide a satisfactory explanation for such optimal properties; instead, one attempts to identify multiple factors, *all* of which are optimized. Meanwhile, "rigorous" formal polynomial lattice models leave open the question of the microscopic mechanisms responsible for the gaps, their inhomogeneities, and the origin of HTSC itself, and even the origin of the HTSC intermediate phase [1,6].

Many readers have found the topological approach described here to be too abstract: they long for some kind of simple analytic model that contains adjustable parameters that can be fitted to experiment. There is an interesting historical precedent for the shift from easily parameterized analytic models to topological models. After Newton solved the problem of planetary motion (essentially a one-body problem in the central field approximation), D'Alembert addressed the hydrodynamic problem of fluid motion. From the Newtonian viewpoint this is a hopelessly complex many-body problem, and even today no exact analytic solutions are known for the general problem of fluid motion subject to general boundary conditions. However, everyone knows that this motion exhibits general properties (turbulent and non-turbulent flow, streaming, eddies, growth rates of unstable modes, etc.). The analysis of these properties has evolved over the last 300 years (not an easy problem!), and topological methods have played a crucial role in that analysis. Conversely, historians say that the calculus of variations originated with D'Alembert's interesting hydrodynamics. This in turn led to the development of Lagrangian mechanics, which contains both analytic and topological elements. The Lagrangian methods involve (topological) paths in configuration space whose properties are constrained by variational principles. Even in vacuum the Lagrangian path approach is useful (Feynman path integrals, for example), but it is much more useful in strongly disordered cases where only large-scale features of the internal atomic structure are known.

Readers who still find this description too abstract should recall three classic problems in mechanics: the disc rolling down an inclined plane without slipping; Huygens' tautochrome (a pendulum whose frequency is independent of amplitude), and Bermoulli's brachistochrome, the curved path between two points with the shortest transit time subject to a constant vertical force. These three apparently different problems all share a common solution (the cycloid), because they contain common variational features, and involve constrained combinations of lateral and vertical motion. In fact, the optimized dopant-centered zigzag current paths in the cuprates all exhibit similar special properties because they also combine constrained lateral and vertical motion!

**2. A Simple (Yet Paradoxical) Example**

The superconductive energy gap $\Delta$ plays a key role in all microscopic theories, starting with the simple metallic "low temperature" superconductors such as Al, Sn and Pb [7]. It is caused by the formation of Cooper pairs, combined in the BCS product many-pair wave function. The gap occurs because the attractive electron-phonon interactions exceed the repulsive Coulomb interaction in the s-wave ($l = 0$) channel. In superfluid $^3$He the $l = 0$ s-wave interaction is repulsive, but the residual $l = 1$ p-wave interaction is weakly attractive, leading to superfluidity at a much lower temperature than in $^4$He. By this reasoning, the appearance of $l = 2$ d-wave anisotropy (with little or no $l = 0$ term) in the energy gap of the cuprates (in early surface measurements of relative grain boundary phases [8], ARPES [9], or Fourier-transformed STM gap patterns [10]) would be consistent only with ultra-low $T_c$'s, so there is a major puzzle here. One way to resolve this puzzle is to note that the d-wave pattern can be strongly enhanced at the surface, and much smaller in the bulk ("d-outside and s-inside") [11]. However, there is a similar, but more fundamental reason, why the d-wave symmetry (which also appears in the pseudogap [12]) is incidental to the microscopic mechanism responsible for HTSC.

The filamentary network model [1] resolves the s-d puzzle in a simple way, which is consistent with the large energy gaps and high $T_c$'s extending up to and including the last surface layer seen in STM experiments. The d-wave anisotropy is exhibited only in the xy plane; using z as the polar axis, this corresponds to an l = 2, m = 2 spherical harmonic. The filamentary model (see Fig. 1) dates back to 1989 [1]:

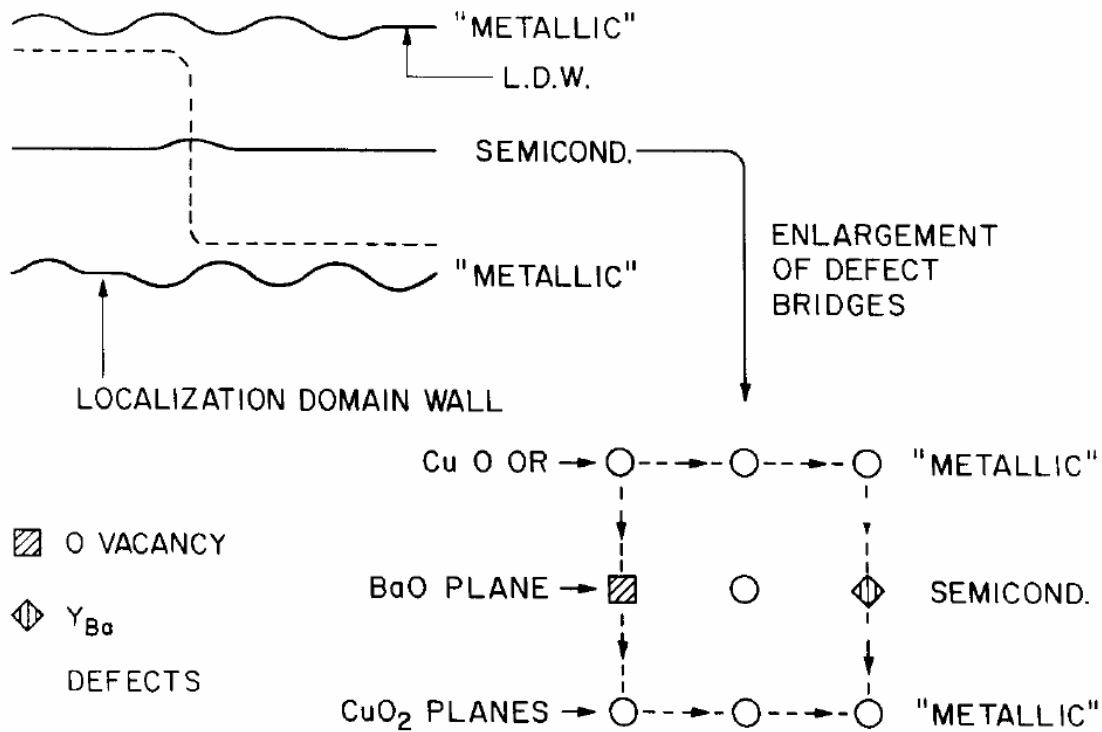

Fig. 1. The basic idea of the filamentary paths in the quantum percolative model [14] for YBCO. The positions of the Insulating Nanodomain Walls (INW pseudogaps) in the *CuO$_2$ layers are indicated, together with the Resonating Tunneling Centers (RTC dopants) in the semiconductive layer, and oxygen vacancies in the CuO$_{1-x}$ chains. Giant e-p interactions are associated with the RTC, where the interactions with LO c-axis phonons are especially large. The INW are perovskite-specific. The sharp bends in the filamentary paths are responsible for the broken symmetry that admixes ab planar background currents with c-axis LO phonons.*

According to the filamentary model, the strong s-wave atomic-scale electron-phonon interactions occur at the interlayer the strong s-wave atomic-scale electron-phonon interactions occur at the interlayer dopants (often interstitial O) that connect conductive layers. At these dopants the path is parallel to the z axis, and the strong s-wave electron-phonon interactions at the interlayer dopants have no effect on the phase coherence of paths projected on the xy plane. There the topologically constrained superconductive paths follow the directions of largest electron wave packet group velocities and strongest local electron-phonon interactions between metal atoms and O atoms, where the longitudinal optic (LO) phonons show strong (10) zone-boundary anomalies (Fig. 2(b)). Thus because the paths are not indexed by crystal momentum, but rather by sets of nearest neighbor vectors in real space, they appear to have d-wave symmetry in the xy plane, while actually representing Cooper pairs formed by s-wave interactions at the dopants. The planar d-wave symmetry is incidental, and is merely the consequence of projecting the zig-zag paths in real space onto xy planar Fourier transforms. Note that these paths must percolate through a planar pseudogap maze that has largely d-wave symmetry, as the LO phonon buckling itself has d-wave symmetry. One technical comment: the s-wave interactions occur on the atomic scale, and technically speaking represent a very large local field correction to the interactions forming Cooper pairs. In the filamentary model they are hidden from, and do not appear in, the x-y planar angular Δ, because of their orthogonal out-of-plane z-axis character.

3. **Elements of the topological model**
The modern theory of glasses is called constraint theory [1]; it describes accurately the phase diagrams of both molecular glasses (window glass) and electronic glasses (the cuprates). (The theory identifies self-consistent network properties, and this accounts for its generic success.) In the molecular case (network glasses such as window glass) the constraints are counted by analyzing interatomic bonding forces (stretching, bending…). The mean-field condition for forming an ideal glass is $N_c = N_d$, where $N_c$ is the number of constraints per atom, and $N_d$ is the number of degrees of freedom per atom. In a gas of N atoms, $N_d = 2d$, but in a glass $N_d = d$, as the glass is frozen into a configuration of nearly maximal density (ideal space filling for a strongly disordered network), with only spatial degrees of freedom (periodic crystalline configurations are, of course, singular and not glassy). Cuprates are electronic glasses, with the dopants frozen into configurations that nearly optimally screen internal electric fields. Constraint theory is topological, not analytical, and it is capable of describing such variational effects by Lagrangian methods without detailed knowledge of glassy configurations, in both molecular (window glass) and electronic (cuprate) glasses. Analytical models of continuum systems have been able to derive results not only for simple nearly free electron metals, but also even for weakly disordered metals (dirty transition metals, …). They are, however, completely unsuited to strongly disordered, nanoscopically inhomogeneous metals with anomalous properties obtained by adaptively doping an insulator, such as the cuprate high temperature superconductors [3-5].

While there are many crystalline materials, there are only a few ideal glass formers. These in turn have properties that are very, very different from those of normal materials, and these properties are often of great value, which has focused attention on them. Constraint theory correspondingly is of little value in discussing most materials, because the ideal glass-forming condition $N_c = N_d$ is seldom satisfied. However, when circumstances make it possible to form ideal glasses, constraint theory is very powerful [1]. It is able to explain, without using adjustable parameters, why cuprates (and to a lesser extent the $(Ba,K)(Pb,Bi)O_3$ family) are the only materials that form high temperature superconductors [12], a highly material-specific result inaccessible to analytic theories, even with the customary adjustable parameters.

Fig. 1 illustrates the basic topological model for layered cuprate high temperature superconductors. Semiconductive layers (such as SrO) alternate with metallic layers ($CuO_2$, BiO). The planar lattice constant is fixed by the isostatically (stress-free) rigid $CuO_2$ layers, which stabilize the structure mechanically and satisfy the ideal glass-forming condition $N_c = N_d$ with regard to interatomic forces [11]. The remaining layers are much softer (or floppier) and are easily distorted: this explains why the only electronic states near $E_F$ that are observable by ARPES are those associated with $CuO_2$ layers. It is customary to assume that the $CuO_2$ layers are perfectly crystalline, but because of the interlayer misfit to semiconductive layers of effectively different relaxed lattice constants, this is almost certainly not the case [1,12,13]. The misfit can be relieved, and the total energy reduced, by inserting thin (semi-)insulating domain walls in the $CuO_2$ layers. Because the Cu d ($x^2 - y^2$)-Op(x,y) conduction band is nearly half-full, the natural choice for such walls would be doubled unit cells, Jahn-Teller distorted in such a way as to introduce a small anisotropic energy pseudogap (or depression in the local density of states), analogous to that formed by a charge density wave. Because of the overall softness of the lattice, the gain in electronic energy from forming such a pseudogap is reduced only partially by the increased elastic strain energy.

The splitting of the metallic planes into nanodomains separated by thin insulating domain walls results in a narrow gap semiconductor. Now as the x dopants are added, usually to the alternating semiconductive layers (such as SrO), they provide bridges that can carry current in zigzag paths around the domain walls (Fig. 1). These paths form dopant-centered conducting wires, and there is an insulator-metal transition when there are enough dopants that the wires begin to percolate at $x = x_1$. This percolative metallic phase is qualitatively different from a quasi-one-dimensional normal metal with weak localization, because the filamentary density increases with x. At $x = x_0$ it reaches a maximum (space is filled), and at larger values of x the filaments bunch together so closely that interfilament scattering converts the bunched regions into regions that are locally normal metals [14]. In these normally metallic regions the local density of states near $E_F$ is much larger than when averaged over the filaments and semicondcutive background. These zigzag paths are a useful tool for analyzing the relation between anomalies in LO phonon spectra measured by neutron scattering [14]. They also provide a derivation [15] of the d-wave gap by projection of the paths on preferred antinodal directions; in this microscopic model nodal gap quasiparticles are replaced by *projections* of Cooper pairs bound to independent filaments onto nodal $(\pi,\pi)$ gap directions; these are the directions along which percolation between nanodomains with edges parallel to Cartesian directions are most effective in dielectric screening.

A more formal description of the origin of the zigzag paths proceeds as follows. For simplicity we assume that all the dopant bridges are locally equivalent. (This is a good approximation because the domain walls are similar; they relieve the global interlayer misfit. Similarly, the dopants are similar, because they adopt the optimal bridging position that maximizes the dopant-assisted tunneling through (or around) the dopant walls.) In a mean field approximation the coherence of the dopant impurity band network is now measured by the structure factor $S(\mathbf{k}) = \Sigma_i \exp(i\mathbf{k}\cdot\mathbf{r}_i)$, where the sum extends over all dopant sites $\mathbf{r}_i$. Such a sum, in the presence of many-electron thermal fluctuations, rapidly becomes incoherent. However, if we separate the sum into its filamenatary components f as a double sum $S(\mathbf{k}) = \Sigma_f \Sigma_i \exp(i\mathbf{k}\cdot\mathbf{r}_{if})$, then each filament contains only a few electrons, so that the average interlevel energy spacing on a given filament (now of order $N^{-1/3}$ instead of $N^{-1}$) becomes large compared to kT. Thus the filaments are separately coherent or incoherent (each filament has its own phase or order parameter), and the coherence of the "intact" filaments is affected only marginally by interfilamentary interactions with the incoherent filaments.

## 4. Soft Host Lattices and Chemical Trends in $T_c^{max}$

In the older metallic and intermetallic superconductors, there were many correlations between lattice softening and higher $T_c$ 's (for instance, the Al lattice is hard, Sn is softer, and Pb softer still, and the $T_c$ 's increase correspondingly). The reason for this trend is that stronger electron-phonon interactions mean better electron-ion dielectric screening of ion-ion repulsion. Are there any corresponding trends in $T_c$, or $T_c^{max}$ (the largest value of $T_c$ for a given alloy family, at P = 0) in the cuprates?

There are. Uemura [16] connected $T_{cs}^{max}$ (optimal doping) and $n_s$, where $n_s$ is estimated from magnetic field penetration depths λ measured by muon spin relaxation. The data show that $T_{cs}^{max}$ is nearly linear in $n_s$ for many cuprates, including samples with $T_c$ depressed by Zn doping. Within the filamentary model $T_c^{max}$ can be estimated as follows. Each filament, with the dopants arranged like pearls on a string, binds its own set of dopant-derived one-electron states that are phase-correlated to produce maximum conductivity, and hence maximum screening of fluctuating internal ionic fields, with the dopants occupying optimized curvilinear threading positions during sample annealing, and refining these positions with decreasing temperature. Below $T_c^{max}$ the fraction of filaments with mutually correlated phases is proportional to $n_s(T)/n_s(0)$. At $T = T_c^{max}$ Cooper pair phase coherence is erased by phase-disruptive interfilamentary phonon absorption. The average spacing between planar filaments (or by three-dimensional filaments projected onto metallic planes) is d, and $dn_s \sim 1$. Thus as $n_s$ decreases, the spacing between paired filaments increases, and $T_c^{max}$ decreases with the absorbed phonon energy.

We now look for the phonon that will be most effective in displacing paired filaments from their optimal configuration. The filaments zigzag from grey dopants outside $CuO_2$ planes (where electron-phonon interactions are large, and $\Delta_s$ is large locally) throughhs in black $CuO_2$ planes (where $\Delta_s$ is small and probably would be 0 except for proximity effects) to the next grey dopant. The parts of the filamentary path most easily disrupted are therefore the weak links in the $CuO_2$ planes, which are common to all the cuprates, and the phonon we are looking for belongs to these planes. Because the $CuO_2$ planes are the stiffest and least disordered element in the host lattice, the dispersion of these phonons is easily measured [17]: the maximum energy at the (100) longitudinal acoustic phonon at the zone boundary is $\omega_0 = 10$ meV. The actual energy of the phase-breaking phonon should be of order $T_c^{max}$. Combining these equations, one finds $T_c^{max} \sim pn_s(0)$. Thus (100) LA phonons set the overall energy scale for HTSC. The proportionality factor p is not easily estimated, as so little is known about filamentary geometry, but it is approximately independent of doping. In particular, even in underdoped samples, the filaments are broadened by phonon-induced proximity effects approaching optimal doping as the interfilamentary barrier $\Delta_p - \Delta_s \rightarrow 0$ with $z \rightarrow 1$. It is this "non-crossing" broadening due to phonons in the thermal bath that makes $T_s(z)$ flatten and appear to be quadratic while $T_p(z)$ remains linear as z increases to 1. On the overdoped side the filaments overlap to form Fermi liquid patches [14] whose area increases as z increases above 1. We can safely assume that p is larger than in conventional superconductors because filamentary (one-dimensional) glassy topology takes advantage of self-organization to be more efficient in constructing high-conductivity vortex loops to expel or screen magnetic fields than (three-dimensional) electron gases.

The complexity of strong glassy disorder generally prevents the successful construction of polynomial models of glasses. In their place one usually finds several trends: these trends reflect the combined effects of optimization of properties of most interest, which in HTSC has consisted largely of maximizing $T_c$ (although other properties are also likely to be important for applications). There are two other trends apart from the Uemura correlation, both referring to variations in $T_c^{max}$ with host lattice properties. Both of these trends are connected with host lattice instabilities and softening; these are a characteristic feature of strong electron-phonon coupling, and have already been observed to limit $T_c^{max}$ in alloys of the old intermetallic superconductors involving (for example) NbN and $Nb_3Sn$ [18].

First-principles calculations of $T_c$ based on electron-phonon interactions (EPI) in self-consistent electronic structures with ideal atomic positions are usually quite accurate for "old" superconductors (such as $MgB_2$ [2]), but such calculations for cuprates yield $T_c$'s too small by factors $\sim$ 100 [19]. This failure indicates the breakdown either of conventional EPI, or of the ideal lattice structure, leading to enormously enhanced interactions due to the glassy character of dopant configurations; experiment has amply demonstrated that the latter are present, as they violently disturb cuprate vibrational spectra [14]. Conventional lattice dynamics (even empirical spring constant models) encounters many technical difficulties in the cuprates. Not only is the number of atoms/(unit cell) large, but also the basic structural unit is actually a nanodomain containing ~ $10^3$ vibrational degrees of freedom. Thus statistical methods become important, and because of dopant disorder the relevant statistics are those of glasses, not gases, liquids or crystals. The cuprates are closely related to perovskites (such as $BaTiO_3$), many of which are ferroelectrics, and nearly all of which are marginally stable elastically and strongly disordered, with nanodomains similar to the cuprates (for instance, manganites [20]).

Lattice softening can be calibrated by treating cuprates and perovskites as incipient glasses, subject to the same axiomatic rigidity rules as network glasses [1]. It is important to realize that these rules have been tested exhaustively, over the last 25 years, against a very large data base derived from commercially very successful materials. These rules are simplest and most easily justified for chalcogenide glass alloys composed of atoms of similar size [21], but more general rules have succeeded for oxide glasses, notably window glass [22], which is 74% $SiO_2$ alloyed with 16% $Na_2O$ and 10% CaO. These chemical proportions of window glass, an ideal, globally and locally stress-free network, are partially explained in terms of the average number <R> of *Pauling* resonating valence bonds/atom, with <R> = 2.40 *exactly* [R = |Z|, Z(Na) = 1, Z(Ca) = 2, Z(Si) = 4, and Z(O) = -2]. The results [22] for the intermediate phase of many ideally stress-free binary and ternary chalcogenide alloy network glasses range from <R> = 2.27 to <R> = 2.52; the entire set of ranges is centered on <R> = 2.40.

Given this background, is there a way to understand both ferroelectrics and cuprates? There is [26]: <R> of many ferroelectric perovskites (such as $BaTiO_3$) is 2.40; these perovskites have large energy gaps, and they can be alloyed with isovalent elements ([Pb,Zr] TiO$_3$), but not doped. With decreasing <R> the lattice softens, and we reach the dopably metallic HTSC cuprates. Here we must be careful, as the values of R can be ambiguous for polyvalent elements like Cu and Bi. We can determine R for these elements by demanding that $\Sigma Z = 0$ for the parent insulator. We then find values at optimal doping which span the range from <R> = 1.67 up to 2.24 (Fig. 2(a)), which lies just below the range [2.27,2.52] of stress-free network glasses. Note that $T_c^{max}$ itself reaches its maximum value at <R> = 2.0. Also note that the progression of the cuprates from larger to smaller values of <R> is also that of the chronological order in which good single-crystal samples became available. This explains why the cuprates are so special: growing good crystal is difficult for <R> < 2.40 (the perovskite value), and the difficulty increases with decreasing <R>. This result and the results shown in Fig. 2 are well outside the reach of any analytical model based on a polynomial Hamiltonian, where all the chemical trends are buried in adjustable parameters.

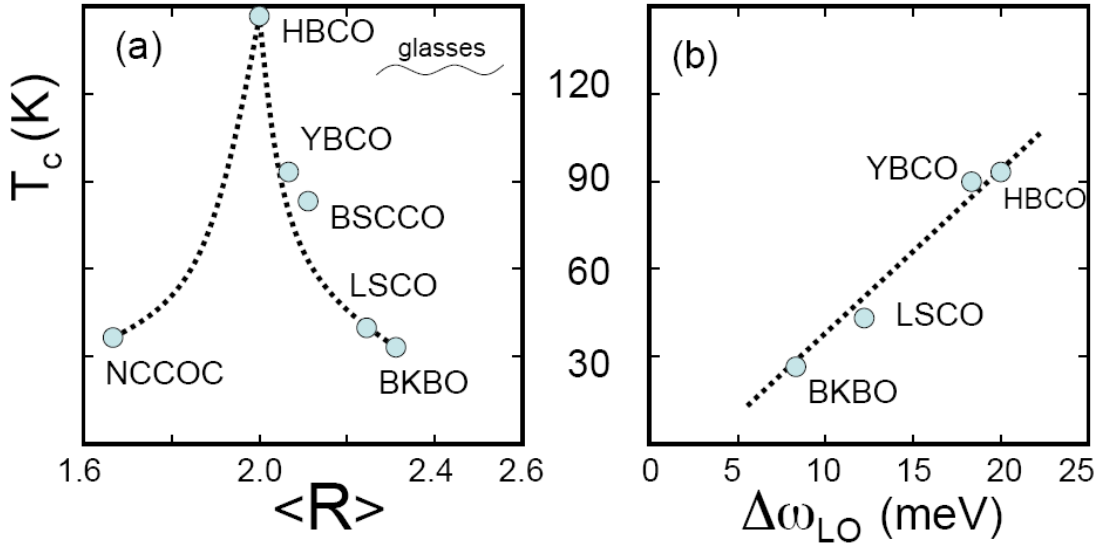

Fig. 2. (a) Chemical trends in $T_c^{max}$ with <R> = $|Z|$, which measures the global stiffness of the doped crystalline network, with Z(Cu) = 2, Z(Bi) = 3 (4) [BSCCO (BKBO), as in the parent insulators, where $\Sigma Z = 0$], and Z(O) = -2. Perovskites (R = 2.40) and pseudoperovskites are only marginally stable mechanically [20], and for HTSC cuprates

*<R> lies in the region of floppy networks just below the isostatic (rigid but unstressed) range determined by studies of network glasses (wavy line). The peak in $T_c^{max}$ occurs at <R> = 2, as one would expect from mean field percolation theory. (b) The cuprates are stabilized by checkerboard reconstruction, the strength of which may determine the $\Delta\omega_{LO}$ phonon ("half-breathing mode") anomaly, which also correlates well with $T_c$. (The single crystal sample [32] of Hg1201 had $T_c$ = 94K.) Dashed lines are guides only.*

In this range oxide crystalline networks are anomalously soft, and one would have expected *a priori* that any metallic states at the Fermi energy should be erased by Jahn-Teller distortions, in the cuprates specifically by buckling of the tetragonal basal planes. Experimentally it has been observed that such buckling is incipient and does limit $T_c$, but the distortions are small because of the isostatic (rigid but unstressed) nature of the $CuO_2$ planes [12]. Thus these rigid planes are not the site of the strong interactions which produce HTSC: quite the opposite, those interactions occur at the dopants in soft planes outside the $CuO_2$ planes, while the planes function in two other ways: (1) as mechanical stabilizers against Jahn-Teller distortions, and (2) as electrical media through which Cooper pairs formed by strong electron-phonon interactions at dopants can connect through S(dopant)-N($CuO_2$ plane)-S(dopant) [SNS] tunneling. Note, by the way, that the oxi-chloride, $Na_xCa_{2-x-y}CuO_2Cl_2$ (NCCOC, $T_c$(x = 0.2, y = 0) = 28K, <R(0.2,0)> = 1.67 ) forms a poor network (because R(Cl) = 1), but with the assistance of Na to bridge the $CuO_2$ planes, self-organization and stabilization by a 4x4 CDW checkerboard [23], it still manages to be superconductive; replacement of Na by Ca vacancies increases <R>, reduces defects, and gives $T_c$(x = 0, y = 0.2) = 38K [24]. Finally, although polaronic effects are strong in all the cuprates [25], they are especially strong in $Na_xCa_{2-x-y}CuO_2Cl_2$ because two ions (Na and Cl) have R = 1.

Perhaps the strongest argument against electron-phonon interactions as the source of HTSC has been the disappearance of the oxygen isotope shift in $T_c$, which is large and normal near the metal-insulator transition (MIT), but decreases towards a small (but still non-zero) value near z = 1 [26]. This long-standing mystery is mitigated by recognizing the variational nature of flexible self-organized percolation. Near the MIT, the paths are far apart and isotopic substitution does not alter the phonon dynamics that causes the normal isotope shift. However, at z = 1 the dynamic effects are compensated by the combined effects of zero-point vibrations [18] and space-filling. Site-selective isotope shifts in the host lattice [26] are an acid test for this explanation. The $CuO_2$ planes are isostatic [12] and nearly ideally crystalline, and hence exhibit an O isotope effect, but the low R planes between them (such as BaO, R = 2) are soft and glassy, so there is no isotope effect at the apical oxygens or the $CuO_x$ chains. This counter-intuitive result is similar to the counter-intuitive clamping (freeing) of the states between $E_F$ and $E_F + \theta_D$ (below $E_F + \theta_D$) observed by ARPES [27], and explained by glassy constraint theory [28].

A third factor involving soft lattices is the local topology associated with the (100) longitudinal optic (LO) phonon kinks [14,29]; these occur near $\mathbf{G}/4 = (1/200)$ and may be related to the 4x4 checkerboard pattern that appears to be associated with pseudogaps in underdoped patches [4]. Probably the most instructive data on the LO phonon anomaly are those [30] taken for YBCO at light doping (x = 0.2) below the MIT, at the MIT (x = 0.35), in the 60K plateau (x = 0.6), and at optimal doping (x = 0.92). Before long CuO chains have formed (x = 0.2) only strong LO scattering (labeled $N_2$) occurs near $\mathbf{q} = (q\ 0\ 0)$ with q = 3.0 at 73 meV, reflecting the strong disorder and the validity of mean-field models. As soon as the CuO chains percolate (x = 0.35), a new LO band (labeled $Z_3$) appears near the zone boundary (q = 3.5) at 57 meV. When the minor cross-linking chains have begun to percolate (x = 0.6), both $N_2$ and $Z_3$ broaden, and the gap $\Delta\omega_{LO}$ near q = 3.25 increases, indicating phase separation between the nanodomains [7] with and without minor chains. At optimal doping (x = 0.92) there is only one phase, but it is ideally glassy [8], and the mean-field component $N_2$ has become very weak, while the percolative component $Z_3$ is very strong.

As all factors must conspire variationally to produce HTSC, one can assume that the LO phonon gap $\Delta\omega_{LO}$ is one of them, and plot $T_c^{max}$ against $\Delta\omega_{LO}$ (Fig. 2(b)): again, there is a strong correlation. Considering that $T_c^{max}$ is limited by $T_p$, and that the pseudogap phase is apparently stabilized by the 4x4 checkerboard pattern associated with it, this correlation is natural. Of course, space filling produces the sharpest LO phonon gap, which occurs at optimal doping ($T_c = T_c^{max}$) in $La_{2-x}Sr_xCuO_4$ (LSCO), x = 0.15; this gap is just as sharp as in the x = 1/8 crystalline "stripe" phase [32], where $T_c = 0$.

One more factor should be considered, and that is the dopant sites themselves. In the presence of multiple nanophases, there may be multiple dopant sites for the same dopant, one for each nanophase. The natural candidate for superconductive interstitial $O_x$ in BSCCO in a superconductive region is a split apical (Cu-O-Bi) site α, which apparently generates polaronic structure in the midinfrared, with a line shape virtually identical to that predicted by theory [31]; α could be associated with a half-filled impurity band pinned to $E_F$, $Z_\alpha = -1$ [32]. The pseudogap regions dominate I-V STM characteristics near $E_F - 0.9$ V, and there a dopant site β is found near Sr [33], $Z_\beta = -2$, that can be assigned to pseudogap regions. Evidence that there is indeed an α site that generates an electrically active impurity band is apparent in the dramatic increase in $N(E_F)$ [34] as T increases across $T_c$. Neutron scattering [35] has provided indirect evidence that there are two independent dopant sites, as those sites are surely responsible for the giant softening of the zone-boundary z-axis polarized phonon observed in metallic LSCO relative to insulating LCO. The phonon peak is anomalously wide, as one would expect if there are two inequivalent Sr doping sites, just as there are two different gap regions.

# REFERENCES

1. J. C. Phillips, Phys. Rev. Lett. **88**, 216401 (2002). Many of the earlier papers in this series can be found by searching SCI with the author's name and the key word "filamentary". A tutorial describing the general properties of self-organized networks, especially the intermediate phase in network glasses, is P. Boolchand, G. Lucovsky, J. C. Phillips, and M. F. Thorpe, Phil. Mag. **85**, 3823 (2005).
2. H. J. Choi, M. L. Cohen, and S. G. Louie, Phys. Rev. B **73**, 104520 (2006).
3. Ch. Renner, B. Revaz, J.-Y. Genoud, K. Kadowaki, and Ø. Fischer, Phys. Rev. Lett. **80**, 149 (1998).
4. T. Cren, D. Roditchev, W. Sacks, J. Klein, J.-B. Moussy, C. Deville-Cavellin, and M. Laguës, Phys. Rev. Lett. **84**, 147 (2000).
5. K. M. Lang, V. Madhavan, J. E. Hoffman, E.W. Hudson, H. Eisaki, S. Uchida and J.C. Davis, Nature **415**, 412 (2002).
6. A. J. Leggett, Nature Physics **2**, 134 (2006**)**
7. J. R.Schrieffer, *Theory of Superconductivity*, (Perseus Books, 1999).
8. C. C. Tsuei, J. R. Kirtley Jr, C. C. Chi *et al*., Phys. Rev. Lett. **73**, 593 (1994).
9. Z.-X. Shen, D. S. Dessau, B. O. Wells, D. M. King, W. E. Spicer, A. J. Arko, D. Marshall, L. W. Lombardo, A. Kapitulnik, P. Dickinson, S. Doniach, J. DiCarlo, T. Loeser, and C. H. Park, Phys. Rev. Lett. **70**, 1553 (1993).
10. K. McElroy, R. W. Simmonds, D.-H. Lee, J. Orenstein, H. Eisaki, S. Uchida, and J. C. Davis, Nature **422**, 592 (2003).
11. K. A. Mueller, Phil. Mag. Letter **82**, 279 (2002).
12. J. C. Phillips, Phil. Mag. **85**, 931 (2005).
13. J. C. Phillips, A. Saxena and A. R. Bishop, Rep. Prog. Phys. **66**, 2111 (2003).
14. J. C. Phillips, Phys. Rev. B **41**, 8968 (1990). J. C. Phillips, Phil. Mag.B **81**, 35 (2001); Phil. Mag.B **82**, 1163 (2002); Phys. Rev. B **71**, 184505 (2005).
15. J. C. Phillips, Phil. Mag. **83**, 3255 (2003); Phil. Mag. **83**, 3267 (2003).
16. Y. J. Uemura, Phys. Rev. Lett. **62**, 2317 (1989); Y. J. Uemura *et al*., Phys. Rev. Lett. **66**, 2665 (1991); Y. J. Uemura , Sol. State Comm. **126**, 23 (2003).

17. P. Böni, J. D. Axe, G. Shirane R. J. Birgeneau, D. R. Gabbe, H. P. Jenssen, M. A. Kastner, C. J. Peters, P. J. Picone, and T. R. Thurston, Phys. Rev. B **38**, 185 (1988).

18. J. C. Phillips, *Physics of High-$T_c$ Superconductivity* (Academic, Boston,1989).

19. K.P. Bohnen, R. Heid, and M. Krauss, Europhys. Lett. **64**, 104 (2003).

20. G. Van Tendeloo, O. I. Lebedev, M. Hervieu, and B. Raveau, Rep. Prog. Phys. **67**, 1315 (2004).

21. S. Chakravarty, D.G. Georgiev, P. Boolchand, and M. Micoulaut, J. Phys. Cond. Mat. **17**, L1 (2005). C. Popov, S. Boycheva, P. Petkov, Y.Nedeva, B. Monchev, and S. Parvanov, Thin Sol. Films **496**, 718 (2006).

22. R. Kerner and J. C. Phillips, Solid State Comm. **117**, 47 (2001).

23. T. Hanaguri, C. Lupien, Y. Kohsaka D.-H. Lee, M. Azuma, M. Takano, H. Takagi, and J. C. Davis, Nature **430**, 1001 (2004).

24. I. Yamada, A. A. Belik, M. Azuma S. Harjo, T. Kamiyama, Y. Shimakawa, and M. Takano, Phys. Rev. B **72**, 224503 (2005).

25. O. Rosch, O. Gunnarsson, X. J. Zhou *et al*., Phys. Rev. Lett **95**, 227002 (2005).

26. H. Keller, Struc. Bonding **114**, 143 (2005); J. Lee, K. Fujita, K. McElroy, *et al*., Nature **442**, 546 (2006).

27. G.-H.. Gweon, T. Sasagawa, S. Y. Zhou, J. Graf, H. Takagi, and A. Lanzara, Nature **430**, 187 (2004).

28. J. C. Phillips, Phys. Rev. B **71**, 184505 (2005).

29. H. Uchiyama, A. Q. R. Baron, S. Tsutsui, Y. Tanaka, W.-Z. Hu, A. Yamamoto, S. Tajima, and Y. Endoh, Phys. Rev. Lett. **92**, 197005 (2004); D. Reznik, L. Pintschovius, M. Ito, S. Iikubo, M. Sato, H. Goka, M. Fujita, K. Yamada, G. D. Gu and J. M. Tranquada, Nature **440**,1170 (2006).

30. Y. Petrov, T. Egami, R. J. McQueeney, M. Yethiraj, H. A. Mook, and F. Dogan, cond-mat/0003414 (2000).

31. J. C. Phillips, Phys. Stat. Sol. **242**, 51 (2005).

32. J. C. Phillips, Proc. Nat. Acad. Sci. (USA) **94**, 12771 (1997); **95**, 7264 (1998).